\documentclass[%
reprint,
 amsmath,amssymb,
 prb,
]{revtex4-2}

\usepackage{latexsym}
\usepackage{graphicx}
\usepackage{times}
\usepackage{color}
\usepackage{dcolumn}
\usepackage{amsmath,amssymb,bm}
\usepackage{here}
\usepackage[FIGTOPCAP]{subfigure}

\begin{document}

\title{Chiral-anomaly-driven magnetotransport in the correlated Weyl magnet Mn$_3$Sn}
\author{S. Kurosawa$^{1}$}
\thanks{shunichiro.kurosawa@nms.phys.s.u-tokyo.ac.jp}
\author{T. Tomita$^{2}$}
\author{M. Ikhlas$^{1,3}$}
\author{M. Fu$^{1,3}$}
\author{A. Sakai$^{1}$}
\author{S. Nakatsuji$^{1,3,4,5,6}$}
\thanks{Author to whom correspondence should be addressed: satoru@phys.s.u-tokyo.ac.jp}

\affiliation{$^1$ Department of Physics, Faculty of Science and Graduate School of Science, The University of Tokyo, Hongo, bunkyo-ku, Tokyo 113-0033, Japan \\
$^2$ National Institute of Advanced Industrial Science and Technology (AIST), Tsukuba, Ibaraki 305-8586, Japan\\
$^3$ Institute for Solid State Physics, University of Tokyo, Kashiwa, Chiba 277-8581, Japan \\
$^4$CREST, Japan Science and Technology Agency (JST), 4-1-8 Honcho Kawaguchi, Saitama 332-0012, Japan \\
$^5$ Institute for Quantum Matter and Department of Physics and Astronomy, Johns Hopkins University, Baltimore, MD 21218, U.S.A \\
$^6$ Trans-scale Quantum Science Institute, University of Tokyo, Hongo, bunkyo-ku, Tokyo 113-0033, Japan}
\date{\today}

\begin{abstract}
The quest for topological states in strongly correlated materials is challenging but essential from fundamental research and application perspectives. The magnetic Weyl semimetal (WSM) state in the chiral antiferromagnet Mn$_3$Sn emerges with strong electronic correlations, offering an intriguing arena for exploring the interplay between Weyl fermions and correlation physics. One prominent characteristic of the WSM state is the chiral anomaly, yet the potential effects of electronic correlations on the chiral anomaly remain unexplored.
Here, we report a comprehensive study of the in-plane magnetotransport properties of single-crystal Mn$_{3+x}$Sn$_{1-x}$ with three different Mn doping levels ($x=0.053$, 0.070, and 0.090). 
The excess Mn leads to glassy ferromagnetic behavior and the Kondo effect, aside from shifting the chemical potential relative to the Weyl nodes. Thus, systematic tuning of the Mn doping level enables us to study the interplay between the spin-fluctuation scatterings, the correlation effect, and the chiral anomaly.
We identify negative longitudinal magnetoresistance and planar Hall effect specific to the chiral anomaly for all three doping levels, indicating that the chiral anomaly persists in the presence of strong correlations.
\end{abstract}
\maketitle
The convergence of topology and correlation physics opens exciting new frontiers in quantum materials research, with magnetic topological semimetal phases serving as a profound example. Weyl fermions, the ever-elusive elementary particles \cite{Weyl1928}, have been found as low-energy quasiparticle excitations in material systems \cite{Xu2015,lv2015experimental,weyl,armitage_RevModPhys.90.015001,nakatsuji_annurev-conmatphys-031620-103859}. In the corresponding Weyl semimetal state (WSM), the Weyl fermions appear at linear crossings of non-degenerate energy bands in momentum space, manifesting topological properties through strongly enhanced Berry curvature inherent to the band touching points. The WSM states occur with either broken inversion or time-reversal symmetry (TRS). The latter case hosted by magnetic materials yield an excellent ground to study the interplay of nontrivial band topology with correlations and the resultant exotic electronic properties ~\cite{armitage_RevModPhys.90.015001,nakatsuji_annurev-conmatphys-031620-103859,nakatsuji_annurev-conmatphys-031620-103859,ahe,kiyohara,machida_2010,XWan_PhysRevB.83.205101,weyl,ELiu_2018,Ye_2018,Ilya_2019,DFLiu_2019,TbMn6Sn6,FeSn,sakai2018giant,sakai_Fe3binary}. Moreover, magnetic WSMs enable the realization of record-high transverse transport, rendering them great potential for applications in spintronics and thermoelectric power generation ~\cite{nakatsuji_annurev-conmatphys-031620-103859,tsai2020electrical,Tsai_2021,otani2021domain,ane,sakai2018giant,sakai_Fe3binary,mizuguchi_2019,smejkal2021anomalous}.

\begin{figure}[t]
    \centering
    \includegraphics[width=\columnwidth]{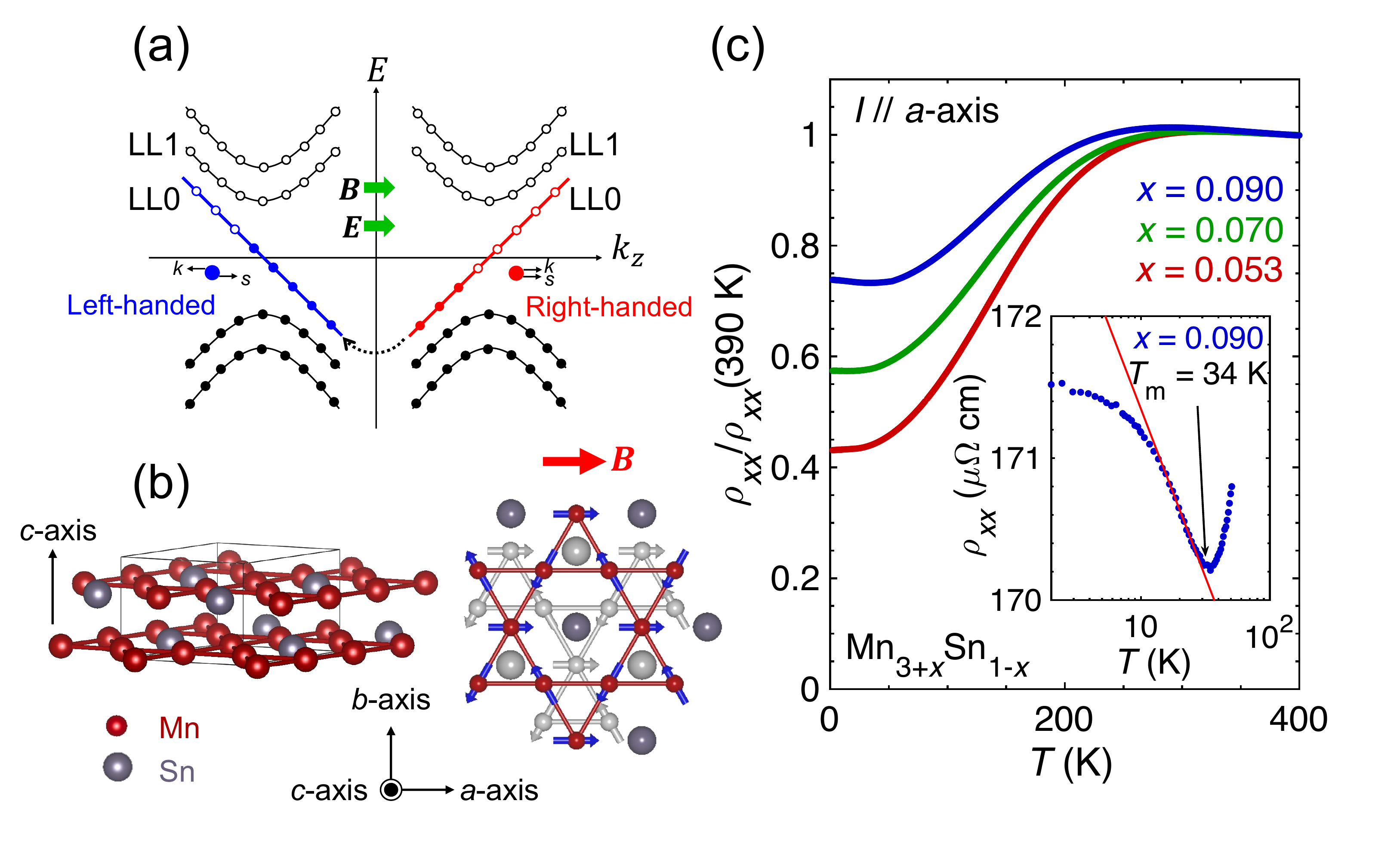}
    \caption{
    (a) Schematics of the Landau levels (LLs) of the Weyl system. The red and blue zeroth LLs (LL0s) correspond to the right- and left-handed chiral levels, respectively. For the applied electric field $E$ parallel to the magnetic field $B$, the charge carriers pump from the right-handed chiral level to the left-handed one, leading to NLMR. (b) Crystal structure and spin configuration of hexagonal Mn$_3$Sn. The left panel shows the side view of the crystal structure, where the Mn kagome lattices stack along $c$-axis ([0001]). The right panel shows the anti-chiral 120$^{\circ}$ order of the Mn moments. (c) Temperature dependence of the normalized zero-field resistivity for the three doping levels measured with current $I$ applied along $a$-axis ($[2\overline{1}~\overline{1}0]$). Inset shows an expanded semi-logarithmic plot of the low-temperature $\rho_{xx}(T)$ for the $x=0.090$ sample. The solid line is a fit to the logarithmic function $\alpha  \ln{T}+\beta$, which yields $\alpha=-1.03$ and $\beta=173.7~\mu\Omega\rm{cm}$.
    }
    \label{sample_evaluation}
\end{figure}

An important hallmark of WSM is the chiral anomaly (Fig. \ref{sample_evaluation}(a)) that leads to defining signatures in magnetotransport: (1) a negative longitudinal magnetoresistance (NLMR) only for parallel electric and magnetic fields ~\cite{nielsen1983adler,son2013chiral,goswami2015axial,hirschberger2016chiral_gdptbi,ong2021experimental}; (2) substantial anisotropic MR (AMR) and planar Hall effect (PHE) under co-planar electric field ${\bf E}$ and magnetic field ${\bf B}$, respectively following $\mathrm{sin2\theta}$ and $\mathrm{cos2\theta}$ dependence on the angle $\theta$ between ${\bf E}$ and ${\bf B}$ ~\cite{nandy2017chiral,burkov2017giant}. These magnetotransport signatures are crucial for identifying WSM states, particularly those coexisting with strong electronic correlations, because correlation effects typically blur the spectroscopic fingerprints of the Weyl nodes in the angle-resolved photoemission spectroscopy (ARPES) measurements. Moreover, theoretical analysis based on first-principle approaches cannot access the detailed characteristics of correlated WSM states; thus, there is a pressing need for experimental studies of the magnetotransport properties.

An exceptional platform to study the interplay between WSM state and electronic correlations is the hexagonal chiral antiferromagnet (AFM) Mn$_3$Sn \cite{weyl}. The crystal structure of this material consists of the Mn kagome lattices stacking along $c$-axis ([0001]) \cite{ahe}, as shown in the left panel of Fig. \ref{sample_evaluation}(b). Below the N\'{e}el temperature $T_{\rm N} \sim 430$ K, the Mn moments form an anti-chiral 120$^{\circ}$ order that breaks TRS and establishes the magnetic WSM state (Fig. \ref{sample_evaluation}(b), right panel) ~\cite{weyl,yang2017topological,chen2021anomalous}. The macroscopic TRS breaking engenders the Weyl nodes and the associated strongly enhanced Berry curvature, leading to surprisingly large anomalous transverse transport and magneto-optical effects comparable to those of conventional ferromagnets, despite the negligible net magnetization of the AFM order ~\cite{ahe, ane, higo2018large_kerr,matsuda2019}. The magnetic Weyl fermions in Mn$_3$Sn are accompanied by strong correlations, as evident from significant bandwidth renormalization and the Kondo effect induced by substantial Mn doping at the Sn sites ~\cite{khadka2020kondo, zhang2020many}. Thus, a systematic study of the chiral-anomaly-driven magnetotransport in Mn$_3$Sn may provide unprecedented insights into intrinsic behavior of the correlated magnetic WSM state.

Here, we carry out an in-depth investigation of the in-plane magnetotransport properties of the single-crystalline Mn$_{3+x}$Sn$_{1-x}$ of three different Mn doping levels, $x=0.053(3),~0.070(4)~\rm{and}~0.090(5)$. The details of experimental methods are presented in the Supplemental Material \cite{supple}. 
In magnetic WSMs, the presence of spin-fluctuation scatterings often complicates the identification of the chiral anomaly. Exploring the evolution of magnetotransport as a function of Mn doping enables us to discern signatures specific to the magnetic WSM from those of spin fluctuations and other conventional mechanisms. Our results show that the chiral anomaly serves as the dominating mechanism for the observed field and angle dependence of magnetotransport in all three samples, indicating that the chiral anomaly of Weyl fermions is robust in the presence of correlation effects.

We first examine the Mn doping effect on the transport properties and magnetization. The normalized zero-field resistivity $\rho_{xx}/\rho_{xx}(390~\mathrm{K})$ of the three Mn$_{3+x}$Sn$_{1-x}$ samples shows qualitatively the same behavior in the anti-chiral 120$^{\circ}$ state for $T>60$ K (Fig. \ref{sample_evaluation}(c)). On further cooling, $\rho_{xx}$ of the $x=0.090$ sample reaches a minimum at $T_{\rm{m}}=34$ K, developing a logarithmic upturn below $T_{\rm{m}}$, and gradually levels off below 10 K (Fig. \ref{sample_evaluation}(c) inset). Such behavior is characteristic of the Kondo effect and is absent for the two lower doping levels $x=0.053$ and $x=0.070$. Our finding is consistent with previous reports that the Kondo effect emerges with substantial Mn doping in single crystals and thin-film samples of Mn$_3$Sn ~\cite{khadka2020kondo, zhang2020many}.

\begin{figure}
    \centering
    \includegraphics[width=\columnwidth]{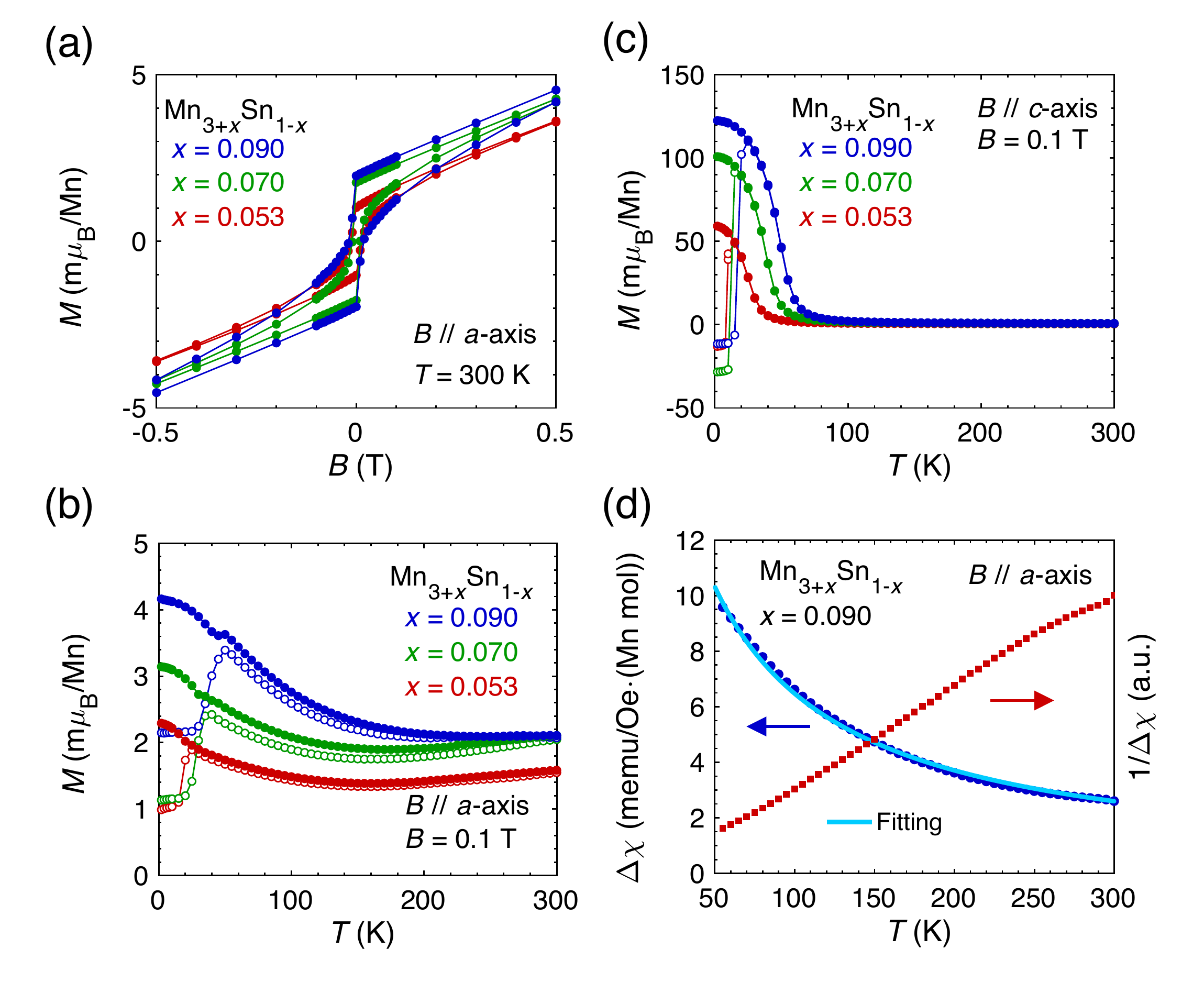}
    \caption{
    (a) Room-temperature field dependence of the magnetization for the three Mn doping levels obtained with $B\parallel a$-axis. 
    (b)(c) Temperature dependence of the magnetization measured under $B=0.1~\rm{T}$ along (b) $a$-axis and (c) $c$-axis. The open and closed circles represent field-cooling (FC) and zero-field-cooling (ZFC) data, respectively. 
    (d) Temperature dependence of the magnetic susceptibility of the $x=0.090$ sample measured with $B\parallel a$-axis. The solid line represents the fitting using the Curie-Weiss law $\Delta\chi = C/(T-\Theta)$, where $\Delta\chi=\chi(x=0.090)-\chi(x=0.053)$, $C$ is the Curie constant and $\Theta$ is the Curie- Weiss temperature. The $1/\Delta\chi$ is represented by the red squares.
    }
    \label{MT}
\end{figure}

To further characterize the behavior of excess Mn moments, we conducted the magnetization measurement, as shown in Figs. 2 (a)--(c). For all samples, we find that the coercivity and the spontaneous magnetization at room temperature are of the order of 100 Oe and a few m$\mu_{B}$/Mn, respectively (Fig. \ref{MT}(a)), typical of Mn$_{3}$Sn \cite{ahe,ane}. Figure \ref{MT}(b) provides the temperature dependence of the magnetization measured under $B=0.1~{\rm T}\parallel a$-axis ($[2\overline{1}$ $\overline{1}0]$). While the magnetization measured along $a$-axis slightly decreases from room temperature to 170~K for the $x= 0.053$ and $x = 0.070$ samples, the magnetization of the $x= 0.090$ sample increases monotonically on cooling from room temperature down to 50~K. We attribute this upturn to the paramagnetic behavior of the excess Mn and carry out Curie-Weiss analysis. 
For fitting Curie-Weiss law $\Delta\chi = C/(T-\Theta)$, where $C$ is the Curie constant, and $\Theta$ is the Curie-Weiss temperature, we subtract the susceptibility of $x=0.053$ sample from one of $x=0.090$ to clarify the paramagnetic behavior of Mn moments: $\Delta\chi=\chi(x=0.090)-\chi(x=0.053)$.
The best fit for $50~\rm{K}<{\it T}<300~\rm{K}$ (the solid line in Fig. \ref{MT}(d)) yields $\Theta=-34$~K and a moment size of $\mu_{\rm{eff}}=2.6~\mu_{\rm B}/{\rm Mn}$ (see Table S1 for details of the fitting parameter values \cite{supple}). 
The size of the moment is rather large given only 0.9~\% extra doping of Mn from $x=0.053$ to $x=0.090$, indicating the cluster formation of the doped Mn moments.
The observed Kondo effect reflects the strong correlations in Mn$_3$Sn system. According to the previous ARPES study combined with DFT analysis \cite{weyl}, the excess Mn induces relative position shift between the chemical potential and the Weyl nodes. Although such an interpretation based on the rigid band picture may hold for the low doping range with itinerant Mn moments, the Kondo physics of localized Mn moments at high doping levels such as $x = 0.090$ should modify the low-energy bands and breaks such a scenario based on the rigid band shift.

Below $T\sim 50~\rm{K}$, the ferromagnetically coupled Mn spin clusters 
lead to a drastic upturn in the magnetization
for the $B\parallel c$-axis (Fig. \ref{MT}(c) and Fig. S2~\cite{supple}
). The in-plane magnetization also displays a relatively mild upturn corresponding to the formation of the clusters (Fig. 2(b)). We determined the onset temperature $T_{\rm p}$ of the ferromagnetic clustering effect by fitting a Gaussian function to the peak in $dM/dT$ vs. $T$ for $B\parallel c$-axis (Fig. S2 inset \cite{supple}). 
With further cooling, spin freezing behavior emerges, as evident from the bifurcation between field-cooling (FC) and zero-field-cooling (ZFC) magnetization curves (Figs. \ref{MT}(b), (c)), and leading to the cluster spin glass (SG) phase. The corresponding transition temperatures ($T_{\rm p}$ and $T_{\rm g}$ along each field direction) are summarized in Table S2~\cite{supple}. 
The excess Mn substituted at the Sn site likely introduces ferromagnetic interaction forming a spin cluster. Thus its frustration against the bulk antiferromagnetic interaction should lead to the formation of SG phase~\cite{glassy,samplequality}.

\begin{figure*}
    \centering
    \includegraphics[width=2\columnwidth]{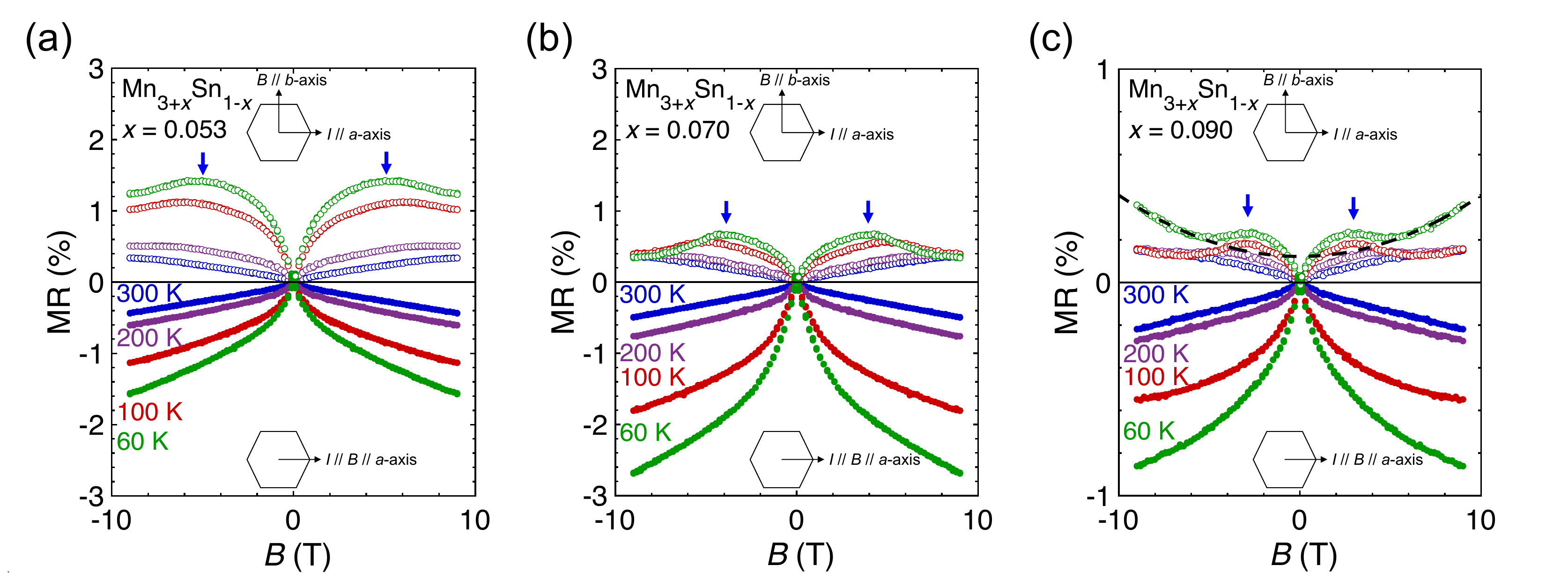}
    \caption{Field dependence of LMR (close symbols) and TMR (open symbols) for (a)~$x=0.053$, (b)~$x=0.070$, and (c)~$x=0.090$ measured at selected temperatures (60 K to 300 K) and under an in-plane magnetic field. Here, the vertical axis represents ${\rm MR (\%)}=100\times(\rho_{xx}(B)-\rho_{xx}(0~{\rm T}))/\rho_{xx}(0~{\rm T})$.
    Blue arrows mark the maximum in TMR at $T=60$~K. The dashed line in (c) is a fit to $B^2$ dependence of TMR at 60 K for the $x=0.090$ sample, which gives ${\rm MR(~\%)}=2.87\times10^{-3}B^{2}+0.122$. The illustrations in each panel show the hexagonal crystal plan of Mn$_3$Sn (See Fig. 1b) and the field $B$ and current $I$ directions (marked by the arrows).
    }
    
    \label{FD_MR}
\end{figure*}
 
 
We next present signatures of the chiral anomaly revealed by in-plane magnetotransport measurements. For all three samples, MR observed in the anti-chiral 120$^\circ$ ordered phase is highly anisotropic: the field-dependent measurements show NLMR for the parallel configuration $B\parallel I\parallel a$-axis whereas positive transverse MR (TMR) for the perpendicular case $B \parallel b$-axis ($[01\overline{1}0]$) and $I \parallel a$-axis (Fig. \ref{FD_MR}). The magnitude of NLMR increases with decreasing temperature from 300 K down to 60 K, consistent with previous studies \cite{weyl}. Experimental artifacts due to current jetting has been ruled out for Mn$_3$Sn \cite{weyl, chen2021anomalous}. Thus, the observed NLMR in Mn$_{3+x}$Sn$_{1-x}$ is intrinsic, unlike those reported in high-carrier-mobility WSM materials \cite{liang2018experimentaltest,ong2021experimental}. Other possible mechanisms for negative MR are the field-induced suppression of spin-fluctuation scattering \cite{yamada1972negative,SMIT1951612} and weak localization \cite{kawaguchi1980negative}. 
Our observation denies both mechanisms; the magnitude of NLMR induced by the former mechanism decreases with decreasing $T$, which contrasts strongly with our observation. The latter mechanism induces negative and dominant MR in all directions and at low field regime \cite{kawabata}, which cannot explain our results. Thus, we can exclude both thermal spin fluctuations and weak localization as the main driver for the observed NLMR; the primary mechanism behind NLMR is the chiral anomaly of magnetic Weyl fermions.
 
Though the overall behavior of the LMR and TMR is qualitatively in line with the chiral anomaly, their variations with Mn doping reveal contribution from the fluctuating Mn spin clusters on top of the chiral-anomaly-induced behavior. On approaching the cluster glass phase ($T\lesssim 100$ K), the positive TMR of all three samples develops a broad, weakly $T$-dependent maximum near 5 T (downward arrows in Fig. \ref{FD_MR}), indicating that a negative MR component emerges. This additional component may arise from the fluctuating Mn defect clusters. At a given $T$, the maximum in TMR shifts to lower $B$ with increasing Mn doping $x$, suggesting that the ferromagnetic coupling among the Mn defect clusters enhances with increasing $x$, leading to the reduced characteristic field for polarizing the spin clusters. Indeed, the negative MR is typically seen in spin glass materials \cite{MR_sg_PhysRevB.71.214424,doi:10.1063/1.333403,Mahendiran_1995}.
The fluctuations of Mn clusters also affect the behavior of NLMR: The LMR is linear-in-field in the single domain state beyond 1 T at high $T$s but develops nonlinear field dependence for $T\lesssim 100$ K, with the curvature change occurring right below the field of the broad maximum in TMR. 
 
We then look into the TMR observed in the $x=0.090$ sample (Fig. \ref{FD_MR}(c)) that shows distinct behavior in different field ranges: steep increase below 3 T, a downturn in the intermediate field range around 5 T, and nearly $B^{2}$ increase at higher fields (the dash line in Fig.\ref{FD_MR}(c)). The low-field increase is likely related to the suppressed carrier hopping due to Zeeman splitting. This mechanism typically occurs when the  bandwidth of electrons exceeds the on-site Coulomb repulsion, resulting in a positive MR that sharply increases in the low field regime and saturates at higher fields \cite{zeeman_theory,zeeman_anderson, zeeman_zno,zeeman_2008}. In the Mn$_{3}$Sn case, the  bandwidth of Mn 3$d$ electrons near the Fermi level is around 4 eV \cite{weyl,stm_Mn3Sn} and the Coulomb repulsion is typically a few eV for $3d$ electrons. Thus, the Zeeman-splitting effect can be relevant for the observed low-field increase in the TMR. The $B^{2}$ behavior at higher fields in the $x=0.090$ sample may arise from orbital MR (OMR) \cite{wte2_li2019anisotropic,mote2,li2018giant_cd3as2,val3}. Then, we roughly estimate the size of the spin fluctuation component at 60 K by removing the $ B^{2}$ OMR background from the total TMR (the dash line in Fig. \ref{FD_MR}(c)). The estimated magnitude of the spin fluctuation component is $0.059~\%$, which is an order of magnitude smaller than the experimental NLMR. This result indicates that the chiral anomaly contribution is still substantial at 60~K and 100~K despite the enhanced fluctuations of the Mn clusters in the highly doped regime.

The doping dependence of the NLMR magnitude reveals correlation effects beyond the rigid band model. At each measured $T$, the magnitude of NLMR first increases with Mn doping from $x=0.053$ to $x=0.070$ then decreases at $x=0.090$, where the excess Mn moments behave as localized magnetic impurities. 
Earlier DFT calculations predict that the increased Mn doping from $x=0.053$ to $x=0.090$ shifts the chemical potential toward the Weyl nodes \cite{ane,weyl,topological_Mn3Sn}, resulting in enhancement of the chiral- anomaly-induced effects. The spin fluctuation contribution should also increase with increasing Mn doping. Thus, the spin fluctuation alone cannot cause the decrease of NLMR magnitude with increasing $x$ from  0.070 to 0.090 if the rigid band model remains valid. Given that the chiral anomaly dominates the NLMR at all doping levels, as discussed above, the observed doping dependence signals the breakdown of the rigid band model for high Mn doping. Namely, above a certain threshold concentration of $x_{\rm c} > 0.070$, the localized Mn moments and the associated Kondo effect likely induce minor modification of the electron bands, more than simply shifting the Fermi level.

 \begin{figure}[htbp]
    \centering
    \includegraphics[width=\columnwidth]{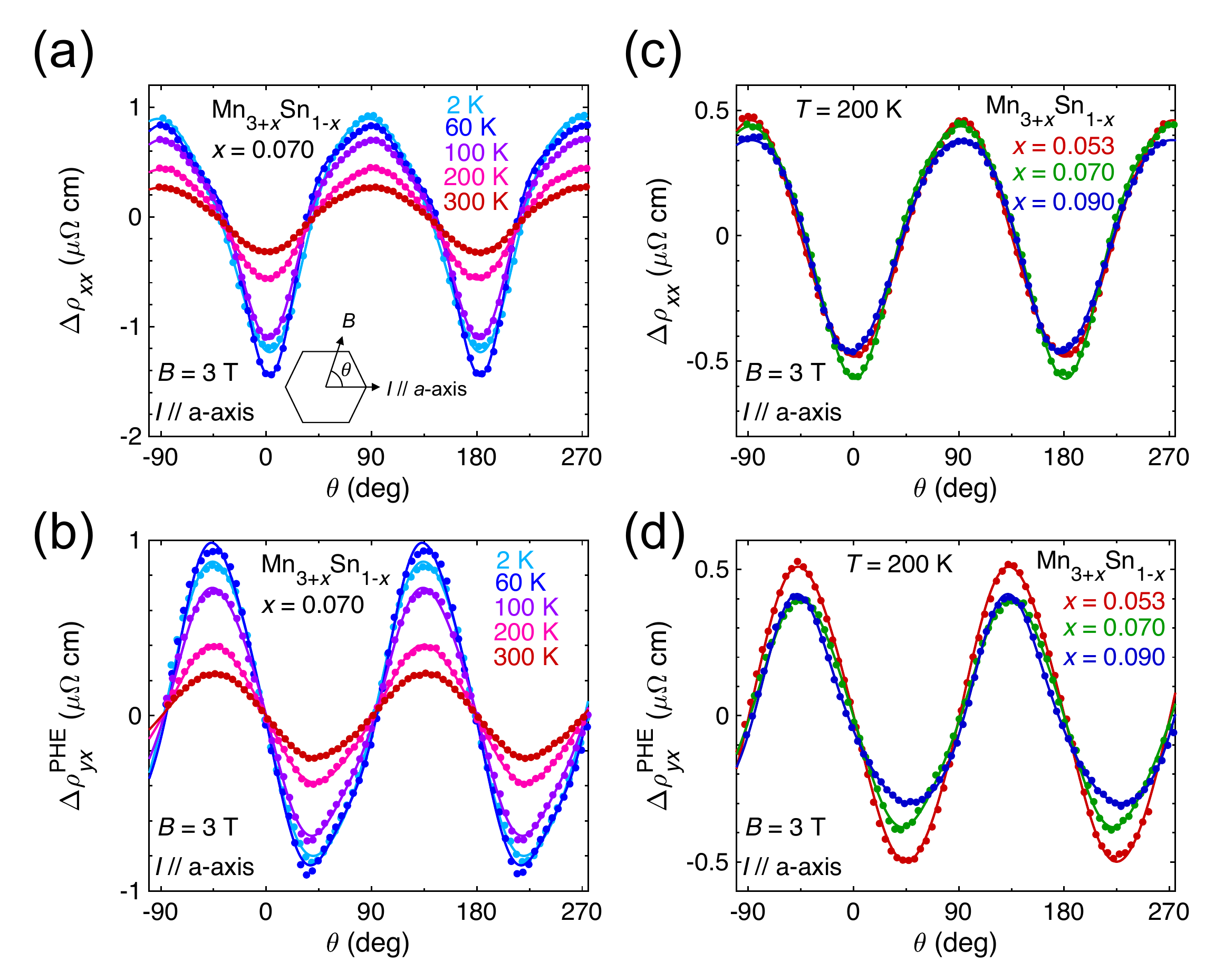}
    \caption{Angular dependence of (a) AMR and (b) PHE of  Mn$_{3+x}$Sn$_{1-x}$ with $x=0.070$ measured at selected temperatures and $B = 3$ T. Inset in (a): The hexagon represents the hexagonal crystal plane of Mn$_3$Sn (see Fig. 1b). The arrows mark the field $B$ and current $I$ directions; $\theta$ is the angle between $B$ and $I$.The comparison of (c) AMR and (d) PHE for three different doping levels at $T=200$ K. The solid lines in all panels are the best fits using Eqs. (\ref{Rxxeqn_6fold}) and (\ref{Rxyeqn_6fold}).
    }
    \label{AD_TD}
\end{figure}

\begin{figure*}[ht]
     \centering
     \includegraphics[width=2\columnwidth]{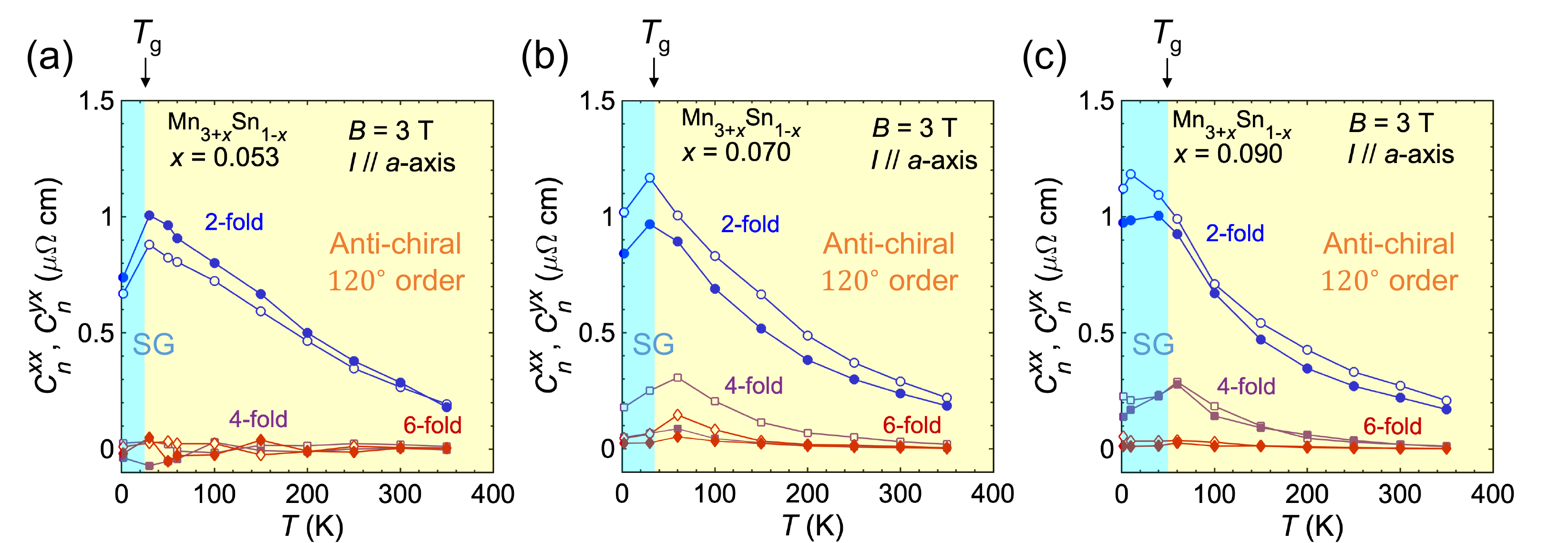}
     \caption{Temperature dependence of the oscillation amplitudes of the angular dependent AMR ($C^{xx}_{n}$, open symbols) and PHE ($C^{yx}_{n}$, closed symbols) of (a)~$x=0.053$, (b)~$x=0.070$, and (c)~$x=0.090$. The amplitudes of two-, four- and six-fold oscillations at 3 T are obtained from the fits with Eqs. (\ref{Rxxeqn_6fold}) and (\ref{Rxyeqn_6fold}). The yellow and blue shades represent the temperature regions of the anti-chiral 120$^\circ$ order and the SG phase, respectively. $T_{\rm{g}}$ represents the SG transition temperature obtained for $B\parallel a$-axis. \cite{supple}.
     }
     \label{amp_TD}
 \end{figure*}
 
To further clarify the contribution from the chiral anomaly, we investigate the angular dependence of the AMR and PHE. We focus on data measured at $B=3$~T, where the single-domain state is established and the influence from field-induced deformation of the spin texture should be negligibly small \cite{li2021free}. Figure \ref{AD_TD} shows the angular dependence of the AMR and PHE as a function of $T$ and $x$, with the current applied along $a$-axis and a rotating field lying in $ab$-plane (Fig. \ref{AD_TD}(a) inset). The oscillating AMR and PHE are predominantly two-fold, with $\Delta\rho_{xx} \propto \cos{2\theta}$ and $\Delta\rho_{yx}^{\rm{PHE}} \propto \sin{2\theta}$, as expected from the chiral anomaly. In several WSMs such as MoTe$_2$~\cite{mote2} and Co$_{3}$Sn$_{2}$S$_{2}$~\cite{co3sn2s2_2020_bulk,co3sn2s2_2020_film}, the LMR is positive, and the observed two-fold oscillations in AMR and PHE mainly arise from the OMR rather than the chiral anomaly. While the OMR may also present in the magnetotransport of Mn$_3$Sn, we observe NLMR in the $I\parallel B$ configuration for the entire measured $T$ range and for all three samples (Fig. \ref{FD_MR}), indicating that the OMR contribution in Mn$_3$Sn is minor and fails to conceal the chiral-anomaly-driven magnetotransport. We note that the small OMR contribution is in line with the short mean free path on the order of $\sim 10~\mathrm{\AA}$, comparable to the lattice parameters. Strong damping of quasi-particles due to electron correlations is observed in ARPES measurements, which should cause the short mean free path as well as the peak in the temperature dependence of the  resistivity near room $T$ (Fig. \ref{sample_evaluation}(c)) \cite{weyl}.

With decreasing temperature (Figs. \ref{AD_TD}(a)(b)) or increasing Mn doping (Figs. \ref{AD_TD}(c)(d)), both AMR and PHE show growing deviations from the two-fold behavior, indicating the presence of higher-order oscillations. A recent theoretical study on Mn$_{3}X$($X$ = Sn, Ge) indicates that the tilted Weyl cones in the $E_{1g}$ magnetic configuration (Fig. \ref{sample_evaluation}(b)) may cause four- and six-fold oscillations \cite{chaudhary2021magnetism}. We then fit the 3 T AMR and PHE data using the equations below:
 \begin{eqnarray}
      \rho_{xx}&=&C_{0}^{xx}+\sum_{\substack{n=2\\n:\rm{even}}}^{6}C_{n}^{xx}\cos(n\theta-\phi_{n}^{xx}) \label{Rxxeqn_6fold}\\
     \rho_{yx}^{\rm{PHE}}&=&C_{0}^{yx}+\sum_{\substack{n=2\\n:\rm{even}}}^{6}C_{n}^{yx}\sin(n\theta-\phi_{n}^{yx})\label{Rxyeqn_6fold}
 \end{eqnarray}
 
The temperature dependence of each component’s amplitude are summarized in Fig. \ref{amp_TD}. For each doping level, the two-fold amplitudes of the AMR and PHE nearly overlap, confirming their identical origin. Specifically, the two-fold oscillations grow monotonically on cooling until peaking at around 10 to 30 K, and then decline with decreasing $T$ in the SG phase. This decline may result from the suppression of the Weyl points in momentum space. In the SG phase, the randomness of the ferromagnetic clusters tends to suppress the Weyl points and thus chiral anomaly because they require the translational symmetry. This is one possible mechanism behind the observed peak near the freezing temperature in the two fold amplitude.

We then discuss the origin of the higher-order component in the observed angular dependence. The amplitudes of the four-fold component is negligible in the entire measured $T$ range in the $x=0.053$ sample (Fig. \ref{amp_TD}(a)) but becomes finite with higher Mn doping (Figs. \ref{amp_TD}(b)(c)). As mentioned above, the increased Mn doping brings the chemical potential closer to the Weyl nodes, so the theoretically predicted higher-order components may be enhanced by Mn doping. This picture can also reasonably explain the more dramatic, non-linear $T$ variation of the two-fold oscillations found at increased $x$ (Figs. \ref{amp_TD}(b)(c)). Moreover, the four-fold component in $x=0.070$ and 0.090 samples peaks at a higher temperature about 60~K than the two-fold components (10 to 30~K). Right below the peak temperature of the four-fold component, the ferromagnetic clusters form, blurring the Weyl points and the associated chiral anomaly. (Fig. S2~\cite{supple}). 
This feature suggests that the four-fold component provides a more sensitive probe to the proximity to the Weyl nodes than the two-fold component. The size of the four-fold component in $x=0.090$ remains substantial; this confirms that the chiral anomaly due to the proximity to the Weyl points is robust against the electron correlation effect.
To summarize, we investigate the properties of in-plane MR using three samples with different Mn doping levels in the Weyl AFM Mn$_3$Sn. From the temperature dependence of the zero-field resistivity and the magnetization, we find that doping of the excess Mn engenders local magnetic moments and the Kondo effect, defying the rigid band picture in the highest Mn doped sample. Unlike other magnetic WSMs, all the samples exhibit NLMR in the anti-chiral 120$^\circ$ order, consistent with the theoretical prediction of the chiral anomaly. The measured AMR and PHE reveal four-fold oscillation in $x=0.070$ and 0.090 samples of Mn$_{3+x}$Sn$_{1-x}$ in addition to the typical two-fold oscillation. The temperature dependence of the oscillatory components of higher doped samples reflect the proximity to the Weyl nodes and the robustness of chiral anomaly against correlation effects. Our comprehensive analyses on chiral-anomaly-induced magnetotransport and magnetism lay the foundation for the studies on WSM in correlated metals.

This work was partially supported by JST-Mirai Program (JPMJMI20A1), JST-CREST (JPMJCR18T3) from Japan Science and Technology Agency, and Grants-in-Aid for Scientific Research (19H00650). The work at the Institute for Quantum Matter, an Energy Frontier Research Center was funded by DOE, Office of Science, Basic Energy Sciences under Award \# DE-SC0019331. 
M.F. acknowledges support from the Japan Society for the Promotion of Science (Postdoctoral Fellowship for Research in Japan (Standard)). The use of the facilities of the Materials Design and Characterisation Laboratory at the Institute for Solid State Physics, the University of Tokyo, is gratefully acknowledged.  

\bibliographystyle{apsrev4-1}
\bibliography{references}

\end{document}